\documentclass[12pt,a4paper,fleqn,leqno]{article} 
\pdfoutput=1

\usepackage{amsfonts}
\usepackage{amsmath,amssymb,amsxtra}
\usepackage{authblk}
\usepackage[english]{babel}
\usepackage{bbm}
\usepackage[font={footnotesize}, skip=5pt, labelfont=bf]{caption}
\usepackage{color}
\usepackage{dsfont}
\usepackage{empheq}
\usepackage{enumitem}
\usepackage{float}
\usepackage[hang, bottom]{footmisc}
\usepackage{geometry}
\usepackage{graphicx}
\usepackage[utf8]{inputenc}
\usepackage{parskip}
\usepackage{ragged2e}
\usepackage{soul}
\usepackage{titlesec}
\usepackage{hyperref}

\setitemize{leftmargin=*, nosep}
\titlespacing{\section}{0mm}{0mm}{1mm}
\titlespacing{\subsection}{0mm}{0mm}{1mm}

\newcommand{\norm}[1]{\lVert #1 \rVert}
\newcommand{\qed}{\hfill\blacksquare}
\newcommand{\R}{\mathbb R}
\newcommand{\roi}{\mathbb R^{m}_{+}\times \mathbb R^{n}_{+}}
\newcommand{\rp}{\mathbb R^{m+n}_{+}}
\newcommand{\ro}{\mathbb R^{m}_{+}}
\newcommand{\ri}{\mathbb R^{n}_{+}}
\newcommand{\sbt}{\,\begin{picture}(-1,1)(-1,-3)\circle*{3}\end{picture}\ }
\newcommand{\trans}[1]{\ensuremath{#1^{\scriptscriptstyle \mathsf{T}}}}
\newcommand{\oset}[1]{\overset{\kern 2pt \lower 2pt \hbox{$\scriptstyle\rightarrow$}}{#1}}

\makeatletter
\def\@biblabel#1{\hspace*{-\labelsep}}
\makeatother

\title{Revisiting Transformation and Directional Technology Distance Functions}
\author{Yaryna Kolomiytseva}
\affil{\small Faculty of Management, Economics and Social Sciences \\ University of Cologne \\ Albertus-Magnus-Platz, 50923 Cologne, Germany \\ Email: kolomiytseva@wiso.uni-koeln.de}
\date{\vspace{-2mm} \small 24 December 2018}

\begin{document}
	\maketitle
	\begin{abstract}
	In the first part of the paper, we prove the equivalence of the unsymmetric transformation function and an efficient joint production function (JPF) under strong monotonicity conditions imposed on input and output correspondences.  Monotonicity, continuity, and convexity properties sufficient for a symmetric transformation function to be an efficient JPF are also stated. In the second part, we show that the most frequently used functional form for the directional technology distance function (DTDF)---the quadratic---does not satisfy homogeneity of degree $-1$ in the direction vector. This implies that the quadratic function is not the directional technology distance function. We provide derivation of the DTDF from a symmetric transformation function and show how this approach can be used to obtain functional forms that satisfy both translation property and homogeneity of degree $-1$ in the direction vector if the optimal solution of an underlying optimization problem can be expressed in closed form.
	\end{abstract}

	
	\newpage
	
	\section{Introduction} 

	Transformation and directional distance functions are frequently employed to represent multiple-output production technology. Much theoretical work has been done to define different types of transformation functions (namely, symmetric and unsymmetric) and joint production functions (namely, isoquant, weak efficient, and efficient) and to establish their properties. Samuelson (1966), Diewert (1973), and Jorgenson and Lau (1974) exploit the definition of the unsymmetric transformation function. This definition entails arbitrarily choosing an output with respect to which maximization is performed given a vector of inputs and of the remaining outputs (Diewert, 1973, p.\ 286). Shephard (1970), Bol and Moeschlin (1975), Al-Ayat and Färe (1979), and Färe et al.\ (1985) study the concept of an isoquant joint production function (JPF) and investigate conditions sufficient for its existence. This function assumes the value zero at the output--input bundles belonging to the isoquant of the production possibilities set whenever this function exists; see Section~2.1 and Färe et al.\ (1985, pp.\ 46--47). A weak efficient and an efficient joint production functions, defined by Färe (1983, 1986), possess the identical property with respect to the weak efficient and the efficient subsets of the production possibilities set, respectively.

	The concepts of a symmetric transformation and a joint production functions are sometimes used interchangeably in the literature; see, for instance, Lau (1972, p.\ 281) or Färe~et~al.\ (1985, p.\ 38). Also, the unsymmetric transformation function is typically presumed, similarly to an efficient joint production function, to trace the efficient subset of the production possibilities set satisfying boundedness, convexity, closedness, and strong disposability conditions; see Diewert (1973, p.\ 286). However, no theoretical results are available on such equivalence. Hence, it is necessary to substantiate this property, potentially under stronger conditions imposed on the production possibilities set, or else to prove the non-equivalence of the unsymmetric transformation and an efficient joint production functions under these conditions. This is the main purpose of the first part of this paper.
	
	In particular, in Section 2.1, we augment the conditions imposed on the production possibilities set by Diewert (1973) with weak attainability of outputs (Shephard, 1970) and weak efficient and efficient strict monotonicity of the input and output correspondences (Färe, 1983; Färe et al., 1985) and examine, in order of increasing restrictiveness, their sufficiency for the existence of three joint production functions and the equivalence of the unsymmetric transformation function and each of the JPFs. We also state our main result, asserting that the unsymmetric transformation function and an efficient JPF are equivalent under weak efficient strict monotonicity of the input and output correspondences if it is technically possible to produce positive quantities of all outputs with each input bundle that makes production feasible. We discuss, in addition, two types of production---joint and assorted---and their compatibility with these conditions.
	
	In Section 2.2, we assume, similarly to Hanoch (1970) and Lau (1972), that a symmetric transformation function is strictly decreasing in inputs and strictly increasing in outputs, continuous, and convex and establish the properties of the production possibilities set induced by this function. We also prove that a symmetric transformation function is an efficient joint production function whenever these monotonicity, continuity, and convexity properties hold. 
	
	Directional distance functions have also received considerable attention in the literature since their introduction by Chambers et al.\ (1996, 1998), drawing on the benefit and shortage functions of Luenberger (1992a, 1992b). Recently, several studies have dealt with econometric estimation of the systems of simultaneous equations encompassing the directional technology distance function (DTDF) and the first-order conditions for cost minimization or profit maximization; see, for example, Atkinson and Tsionas (2016) and Malikov et al.\ (2016). In these studies, the most frequently used functional form for the DTDF is a quadratic that is restricted to satisfy translation property. In Section 3.1, we show that the quadratic function does not satisfy homogeneity of degree $-1$ in the direction vector, which implies that this function is not, in fact, the directional technology distance function.
	
	Since functional forms satisfying both translation property and homogeneity of degree $-1$ in the direction vector are not readily available, in Section 3.2, we provide derivation of the directional technology distance function from a symmetric transformation function satisfying the properties analyzed in Section 2.2. We conclude with an example in which the transformation function is separable in inputs and outputs with a quadratic output and a linear input functions. The proposed approach, however, has certain limitations, which preclude the use of those functional forms for a transformation function for which the optimal solution of an underlying optimization problem cannot be expressed in closed form.	
	
	\bigskip
	We use the following notation for vector inequalities:
	\begin{itemize}
		\item[] $x \geqq y$ if and only if $x_i \geq y_i$ for all $i$; 
		\item[] $x \geq y$ if and only if $x \geqq y$ and $x \neq y$;
		\item[] $x \stackrel{\ast}{>} y$ if and only if $x_i > y_i$ or $x_i = y_i = 0$ for all $i$;
		\item[] $x > y$ if and only if $x_i > y_i$ for all $i$.
	\end{itemize} 
	In addition, $\R_+ = \{\mu \in \R\mid \mu \geq 0\}$, $\ro = \{y \in \R^m\mid y \geqq 0\}$, and $\rp = \roi$. Proper inclusion is denoted by the symbol $\subset$.
	\newpage

    \section{Transformation functions}
    \vspace{-2mm}
	\subsection{Unsymmetric transformation function}
	
	Let $(y,x)$ denote a nonnegative output--input bundle, where $y \in \ro$ and $x \in \ri$. We follow Diewert (1973, p.\ 286) in assuming that the production possibilities set $T$ satisfies conditions T1--T5:
	\begin{enumerate}
		\item[T1.] $T$ is a nonempty subset of $\ro\times\ri$; in particular, $(0,0) \in T$;
		\item[T2.] $T$ is closed;
		\item[T3.] $T$ is convex;
		\item[T4.] if $(y, x) \in T$ and $(-y',x') \geqq (-y,x)$, then $(y',x') \in T$;
		\item[T5.] $P(x)$ is bounded for all $x \geq 0$.
	\end{enumerate}
	Condition T1 is here slightly modified to allow for possibility of inaction. In condition~T5, the output correspondence $P\colon \ri \to 2^{\ro}$ is given by $P(x) = \{y \in \ro\mid (y,x) \in T\}$.  
	
	Imposing conditions T1--T5 on the production possibilities set, Diewert (1973, p.\ 287) defines the unsymmetric transformation function $t\colon \R^{m-1}_{+}\times \ri \to \R_{+}\cup\{-\infty\}$ in the following way.
	
	\textbf{Definition 2.1.1 (Diewert, 1973).} For all $(y^{-i}, x) \in \R^{m-1}_{+}\times \ri$,
	\begin{equation*}
	t(y^{-i},x) = \begin{cases} \max\{v_i \in \R_{+}\mid (y_1, \dots, v_i, \dots, y_m) \in P(x)\} & $if $ (y_1, \dots, v_i, \dots, y_m) \in P(x) \\ & $for some $  v_i \in \R_{+}; \\ -\infty & $otherwise$, \end{cases}
	\end{equation*}
	where $y^{-i} = (y_1, \dots, y_{i-1}, y_{i+1}, \dots, y_m)$ and $i \in \{1, \dots, m\}$.
	
	Adjusting the definition of an isoquant joint production function given by Shephard (1970, p.\ 213), Färe (1986, p.\ 672) defines, in addition, a weak efficient and an efficient joint production functions as follows\footnote{The definitions of the isoquant, the weak efficient, and the efficient subsets of an input or output set are given in the Appendix.}. 
	
	\textbf{Definition 2.1.2 (Färe, 1986).} A function $I\colon \roi \to \R$ such that
	\begin{enumerate}
		\item[(i).] for all $x \geq 0$ with $P(x) \neq \{0\}$, Isoq/WEff/Eff $P(x) = \{y \in \ro\mid I(y, x) = 0\}$;
		\item[(ii).] for all $y \geq 0$ with $L(y) \neq \varnothing$, Isoq/WEff/Eff $L(y) = \{x \in \ri\mid I(y, x) = 0\}$
	\end{enumerate}
	is called an isoquant/weak efficient/efficient joint production function.
	\newpage 
	
	Suppose there exist $x \in \ri$ such that $P(x) = \{0\}$ and $y \in \ro$ such that $L(y) = \varnothing$. Define 
	\begin{minipage}{0.49\textwidth}
		\begin{equation*}
		\begin{split}
		& X_1 = \{x \in \ri\mid x \geq 0 \text{ and } P(x) \neq \{0\}\}, \\
		& X_2 = \{x \in \ri\mid x \geq 0 \text{ and } P(x) = \{0\}\}, \\
		& X_3 = \{0\},		
		\end{split}
		\end{equation*}
    \end{minipage}
	\hfill
	\begin{minipage}{0.49\textwidth}
		\begin{equation*}
		\begin{split}
		& Y_1 = \{y \in \ro\mid y \geq 0 \text{ and } L(y) \neq \varnothing\}, \\
		& Y_2 = \{y \in \ro\mid y \geq 0 \text{ and } L(y) = \varnothing\}, \\
		& Y_3 = \{0\}.		
		\end{split}
		\end{equation*}
    \end{minipage}
 
	\medskip
	Then $\{X_1, X_2, X_3\}$ is a partition of $\ri$ and $\{Y_1, Y_2, Y_3\}$ is a partition of $\ro$. Bol and Moeschlin (1975, p.\ 395) prove that an isoquant JPF exists if and only if, for all $(y,x) \in Y_1\times X_1$, we have $x \in \text{Isoq } L(y)$ if and only if $y \in \text{Isoq } P(x)$. Here, $L\colon \ro \to 2^{\ri}$ is the input correspondence, which is the inverse of $P$, given by $L(y) = \{x \in \ri\mid (y,x) \in T\}$. Furthermore, suppose that an isoquant joint production function exists. Then
	\begin{equation*}
	\begin{split}
	& \{(y,x) \in Y_1\times X_1\mid x \in \text{Isoq } L(y) \text{ and } y \in \text{Isoq } P(x)\} = \\
	& \{(y,x) \in Y_1\times X_1\mid x \in \text{Isoq } L(y)\} = \{(y,x) \in Y_1\times X_1 \mid y \in \text{Isoq } P(x)\}, \\
	\end{split}
	\end{equation*}
	and a function $I\colon \roi \to \R$ is an isoquant joint production function if and only if 
	\begin{equation*}\tag{2.1.3}
	\begin{split}
	& \{(y,x) \in (\ro\times X_1)\cup(Y_1\times\ri)\mid I(y,x) = 0\} = \\
	& \{(y,x) \in Y_1\times X_1\mid x \in \text{Isoq } L(y) \text{ and } y \in \text{Isoq } P(x)\}.
	\end{split}
	\end{equation*}
	We will use these results extensively in this section. The value of an isoquant JPF is unrestricted, i.e., it may or may not assume the value zero, if $(y,x) \in (Y_2\cup Y_3)\times(X_2\cup X_3)$. 
	
	\medskip
	\textbf{Existence of isoquant JPF}

	We now show that conditions T1--T5 imposed on the production possibilities set are not sufficient for the existence of an isoquant JPF. Let $P(x) = \{y \in \R_{+}\mid y \leq h(x)\}$ for all $x \in \R_{+}$, where the function $h\colon \R_{+} \to \R_{+}$ is given by\footnote{This example is adapted from Färe et al.\ (1985, pp.\ 31--32).}
	\begin{equation*}
	h(x) = \begin{cases} x & $if $ x \in [0,1); \\ 1 & $if $ x \in [1,\infty). \end{cases}
	\end{equation*}
	In this case, the graph of $P$ satisfies conditions T1--T5; however, $1 \in$ Isoq $P(2)$ and $2 \not\in$ Isoq $L(1)$. Hence, an isoquant joint production function does not exist. 
	
	Consider, in addition, condition T6$^\ast$, weak attainability of outputs, stated by Shephard (1970, p.\ 185):
	\begin{enumerate}
		\item[T6$^\ast$.] if $x \geq 0$, $y \geq 0$, and $y \in P(\lambda x)$ for some $\lambda > 0$, then for each $\theta > 0$ there exists $\lambda_\theta > 0$ such that $\theta y \in P(\lambda_\theta x)$.
	\end{enumerate}
	\textbf{Proposition 2.1.4.} If the production possibilities set satisfies conditions T1--T5 and T6$^\ast$, then an isoquant joint production function exists.
	
	\textsc{Proof.} First, we show, following Bol and Moeschlin (1975, p.\ 397), that $y \in \text{Isoq } P(x)$ implies $x \in \text{Isoq } L(y)$ for all $(y,x) \in Y_1\times X_1$. Suppose to the contrary that there exists $(y,x) \in Y_1\times X_1$ such that $y \in \text{Isoq } P(x)$ and $x \not\in \text{Isoq } L(y)$. Then $\lambda x \in L(y)$ for some $\lambda < 1$. Fix $\theta > 1$. By weak attainability of outputs, there exists $\lambda_\theta > 0$ such that $\theta y \in P(\lambda_\theta x)$. Furthermore, $\lambda_\theta > 1$, since $y \in \text{Isoq } P(x)$ and $\theta > 1$ imply that $\theta y \not\in P(x)$ and, therefore, $\lambda_\theta x \not\in L(\theta y)$ if $\lambda_\theta \leq 1$, by strong disposability of inputs. Since $T$ is convex, $((\tau + (1 - \tau)\theta) y, (\tau\lambda + (1 - \tau)\lambda_\theta) x) \in T$ for all $\tau \in [0,1]$. Also, there exists $\widehat{\tau} \in (0,1)$ such that $\widehat{\tau}\lambda + (1 - \widehat{\tau})\lambda_\theta = 1$. Hence, $(\widehat{\theta}y, x) \in T$ for $\widehat{\theta} = \widehat{\tau} + (1 - \widehat{\tau})\theta > 1$ and, therefore, $y \not\in \text{Isoq } P(x)$, which leads to a contradiction. 
	
	Conversely, suppose $x \in \text{Isoq } L(y)$ and $y \not\in \text{Isoq } P(x)$. Then there exists $\theta > 1$ such that $\theta y \in P(x)$. By convexity of $T$ and possibility of inaction, $(\tau\theta y, \tau x) \in T$ for all $\tau \in [0,1]$. Setting $\tau = 1/\theta$ yields a contradiction. Thus, an isoquant JPF exists whenever the production possibilities set satisfies conditions T1--T5 and T6$^\ast$. $\qed$ 
	
	\medskip
	\textbf{Equivalence of unsymmetric transformation function and weak efficient JPF}
	
	In this subsection, we consider a symmetric representation $F^{i}\colon \roi \to \R\cup\{\infty\}$ of the unsymmetric transformation function $t$ given by $F^{i}(y,x) = y_i - t(y^{-i},x)$ for all $i \in \{1,\dots,m\}$ and examine whether or not the unsymmetric transformation function is equivalent to an isoquant joint production function $I$ under conditions T1--T5 and T6$^\ast$. Somewhat abusing terminology, we say that the functions $t$ and $I$ are equivalent if
	\begin{equation*}\tag{2.1.5}
	\begin{split}
	& \{(y,x)\in (\ro\times X_1)\cup(Y_1\times\ri)\mid I(y,x) = 0\} = \\
	& \{(y,x)\in (\ro\times X_1)\cup(Y_1\times\ri)\mid F^{i}(y,x) = 0\} \text{ for each } i \in \{1, \dots, m\}.
	\end{split}
	\end{equation*}
	Thus, the unsymmetric transformation function and a JPF are equivalent if the subsets upon which a symmetric representation of $t$ and $I$ assume the value zero are equal, regardless of an output with respect to which maximization is performed in Definition 2.1.1. It follows from (2.1.3) that the functions $t$ and $I$ are equivalent if and only if $F^{i}$ is a joint production function for all $i \in \{1, \dots, m\}$.   
	
	In fact, provided that conditions T1--T5 and T6$^\ast$ hold, it is not necessary to consider from now on an isoquant and a weak efficient JPFs separately. Färe et al.\ (1994, pp.\ 40--41) prove that Isoq $P(x) = $ WEff $P(x)$ for all $x \in X_1$ and Isoq $L(y) = $ WEff $L(y)$ for all $y \in Y_1$ whenever inputs and outputs are strongly disposable. However, if conditions T1--T5 and T6$^\ast$ are not sufficient for WEff $P(x) = $ Eff $P(x)$ for all $x \in X_1$, it might occur that $F^{i}(y,x) \neq 0$ for some $(y,x) \in Y_1\times X_1$ with $y \in$ WEff $P(x)$, when $i$ is chosen arbitrarily. The following example demonstrates that conditions T1--T5 and T6$^\ast$ imposed on the production possibilities set do not imply that the weak efficient and efficient subsets of output sets are equal.
	\newpage

	\textbf{Example 2.1.6.} For all $x \in \R_{+}$, let $P(x) = \{y \in \R^2_{+}\mid y_2 \leq x \text{ and } y_1 + y_2 \leq 2x\}$ with \mbox{WEff $P(x) \neq$}  Eff $P(x)$ for all $x > 0$; see Figure 1. Its inverse correspondence is given by $L(y) = \{x \in \R_{+}\mid x \geq \max\{y_2, \frac{1}{2}(y_1 + y_2)\}\}$ and their isoquants by 
	\begin{equation*}
	\text{Isoq } P(x) = \{y \in \R^2_{+} \mid 0 \leq y_1 \leq x \text{ and }  y_2 = x\}\cup \{y \in \R^2_{+} \mid 0 \leq y_2 \leq x \text{ and } y_1 = 2x - y_2\}
	\end{equation*} 
	and Isoq $L(y) = \{\max\{y_2, \frac{1}{2}(y_1 + y_2)\}\}$. Since $x \in$ Isoq $L(y)$ if and only if $y \in$ Isoq $P(x)$ for all $(y,x) \in Y_1\times X_1$, an isoquant joint production function exists. 
	
	\begin{figure}[H]  
	\begin{center}
		\includegraphics[width = 0.7\textwidth,trim={0 0.5cm 1cm 1.5cm},clip]{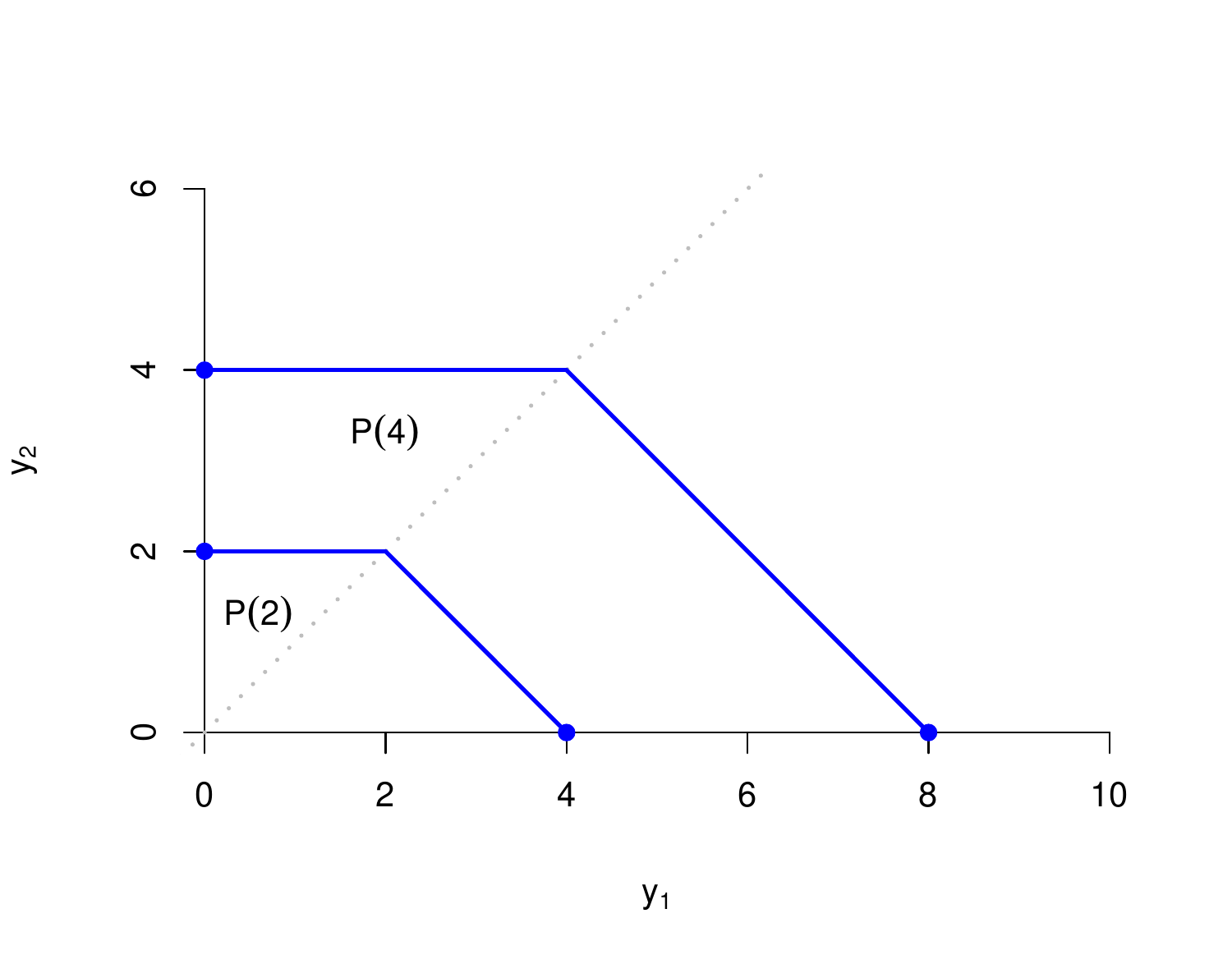}
		\caption{$P(x) = \{y \in \R^2_{+}\mid y_2 \leq x \text{ and }  y_1 + y_2 \leq 2x\}$ for all $x \in \R_{+}$.}
	\end{center}
	\end{figure}
	\vspace{-3mm}
	For the output correspondence $P$, it is easily seen that conditions T4--T5 and T6$^\ast$ hold, so it remains to demonstrate T2 and T3. Shephard (1970, p.\ 300) states that $T$ is convex if and only if $\tau P(x) + (1 - \tau) P(z) \subseteq P(\tau x + (1-\tau)z)$ for all $x, z \in \R_{+}$ and for all $\tau \in [0,1]$. If $y \in P(x)$ and $w \in P(z)$, then $y_2 \leq x$, $y_1 + y_2 \leq 2x$ and $w_2 \leq z$, $w_1 + w_2 \leq 2z$, implying that $\tau y_2 + (1 - \tau) w_2 \leq \tau x + (1 - \tau)z$ and $\tau (y_1 + y_2) + (1 - \tau)(w_1 + w_2) \leq 2(\tau x + (1-\tau)z)$. Hence, $\tau y + (1-\tau)w \in P(\tau x + (1 - \tau)z)$ and the graph of $P$ is convex.
	
	Lastly, we show that $P$ is closed-valued and upper hemicontinuous; condition T2 then follows. A set $P(x)$ is closed for all $x \in \R_{+}$, since it is the intersection of a finite number of closed half-spaces. To see that $P$ is upper hemicontinuous, consider arbitrary sequences $\{x_n\}$ converging to some $x \in \R_{+}$ and $\{y_n\}$ such that $y_n \in P(x_n)$ for all $n \in \mathbb{N}$.  Since $\{x_n\}$ is bounded, the sequence $\{y_n\}$ is also bounded and, thus, contains a subsequence $\{y_{n_k}\}$ that converges to some $y \in \R^2_{+}$. Let $y_{n_k}^i$ denote the $i$th component of a vector $y_{n_k}$. It follows that $y_2 = \lim y_{n_k}^2 \leq \lim x_{n_k} = x$ and $y_1 + y_2 = \lim\big(y_{n_k}^1 + y_{n_k}^2\big) \leq \lim 2x_{n_k} = 2x$ and, therefore, $y \in P(x)$. 
	
	\textbf{Lemma 2.1.7.} Let the production possibilities set satisfy conditions T1--T5 and T6$^\ast$. If WEff $P(x_0) \neq$ Eff $P(x_0)$ for some $x_0 \in X_1$, then $F^{i}$ is not a weak efficient joint production function for some $i \in \{1,\dots,m\}$.
	
	\textsc{Proof.} Suppose there exists $x_0 \in X_1$ such that WEff $P(x_0) \neq$ Eff $P(x_0)$ and choose $y$ in WEff $P(x_0)\setminus$Eff $P(x_0)$. Since $y \not\in$ Eff $P(x_0)$, there exists $v \in P(x_0)$ such that $v_i > y_i$ for at least one $i$ and $v_j \geq y_j$ if $j \neq i$. Let $D_v = \{u \in  \ro\mid u \leqq v\}$. By strong disposability of outputs, $D_v \subseteq P(x_0)$ and also $D_v \ni \bar{u}$ such that $\bar{u}_i = v_i$ and $\bar{u}_j = y_j$ if $j \neq i$. It follows that $t(y^{-i}, x_0) \geq \bar{u}_i > y_i$, implying that $F^{i}(y, x_0) < 0$. We thus conclude that WEff $P(x_0) \not \subseteq \{y \in \ro\mid F^{i}(y, x_0) = 0\}$ for this $i$. $\qed$
	
	In summary, even though conditions T1--T5 and T6$^\ast$ imply the existence of an isoquant and a weak efficient joint production functions, they do not guarantee that, for an arbitrarily chosen $i$, the unsymmetric transformation function would trace the whole weak efficient subset of an output set if the weak efficient and efficient subsets of its boundary are not equal.  
	
	\medskip
	\textbf{Existence of efficient JPF}
	
	In the previous subsection, we established that conditions T1--T5 and T6$^\ast$ are not sufficient for the equivalence of the unsymmetric transformation function $t$ and a weak efficient JPF. However, whether or not they are sufficient for the equivalence of $t$ and an efficient JPF still remains unanswered. In this subsection, we show that conditions T1--T5 and T6$^\ast$ do not imply the existence of an efficient JPF, discuss several monotonicity conditions from Färe (1983) and Färe et al.\ (1985), and prove the equivalence of the unsymmetric transformation function and an efficient joint production function under these conditions.
	
	\textbf{Lemma 2.1.8 (Färe, 1983).} An efficient joint production function exists if and only if, for all $(y,x) \in Y_1\times X_1$, we have $y \in \text{Eff } P(x)$ if and only if $x \in \text{Eff } L(y)$.
	
	Example 2.1.6 demonstrates that conditions T1--T5 and T6$^\ast$ do not imply the existence of an efficient JPF. Therein, the output correspondence $P$ satisfies conditions T1--T5 and T6$^\ast$; however, $(1/2,1) \not\in$ Eff $P(1)$ and $1 \in$ Eff $L(1/2,1)$. Therefore, an efficient joint production function does not exist. 
	
	Färe (1983, pp.\ 16--17) introduces additional condition T7$^\ast$, termed efficient strict monotonicity of the input and output correspondences, and proves its necessity and sufficiency for the existence of an efficient joint production function whenever inputs and outputs are strongly disposable: 
	\begin{enumerate}
		\item[T7$^\ast$.] for all $y \in Y_1$ and for all $x \in X_1$,
			\begin{equation*}
			\begin{split}
			\text{E1. } & \text{Eff } L(y) \cap \text{Eff } L(v) = \varnothing \ \text{ if } \ y \geq v; \\
			\text{E2. } &\text{Eff } P(x) \cap \text{Eff } P(z) = \varnothing \ \text{ if } \ x \geq z.
			\end{split}
			\end{equation*}
	\end{enumerate} 
	\vspace{-2mm}
	Intuitively, E1 states that if the feasibility of a production plan is not retained when any input is decreased, then an increase in any output is also not feasible. E2, in turn, states that if an increase in any output is not feasible, then a decrease in any input is also not feasible; see also Färe (1983, p.\ 16). 
	
	\medskip
	\textbf{Equivalence of unsymmetric transformation function and efficient JPF}
	
	Although conditions T1--T5 and T6$^\ast$--T7$^\ast$ imply the existence of an efficient joint production function, their sufficiency for the equivalence of the unsymmetric transformation function and an efficient JPF is not readily seen. In fact, if these conditions do not rule out the existence of $x \in X_1$ such that WEff $P(x) \neq$ Eff $P(x)$, it might occur that $F^{i}(y,x) = 0$ for some $(y,x) \in Y_1\times X_1$ with $y$ belonging to the weak efficient but not to the efficient subset of $P(x)$, when $i$ is chosen arbitrarily. The following example and lemma demonstrate that this is indeed the case.
	
	\textbf{Example 2.1.9.}  For all $x \in \R^2_{+}$, let $P(x) = \{y \in \R^2_{+}\mid y_2 \leq x_2 \text{ and } y_1 + y_2 \leq x_1 + x_2\}$; see Figure~2. If one of the components of $x$ is zero, then WEff $P(x) = $ Eff $P(x)$; however,  WEff $P(x) \neq $ Eff $P(x)$ if $x > 0$.  This output correspondence satisfies conditions T1--T5 and T6$^\ast$; an argument similar to the one used in Example 2.1.6 shows that the graph of $P$ is convex.  
	\vspace{-3mm}
	\begin{figure}[H]  
	\begin{center}
		\includegraphics[width = 0.7\textwidth,trim={0 0.5cm 1cm 1.5cm},clip]{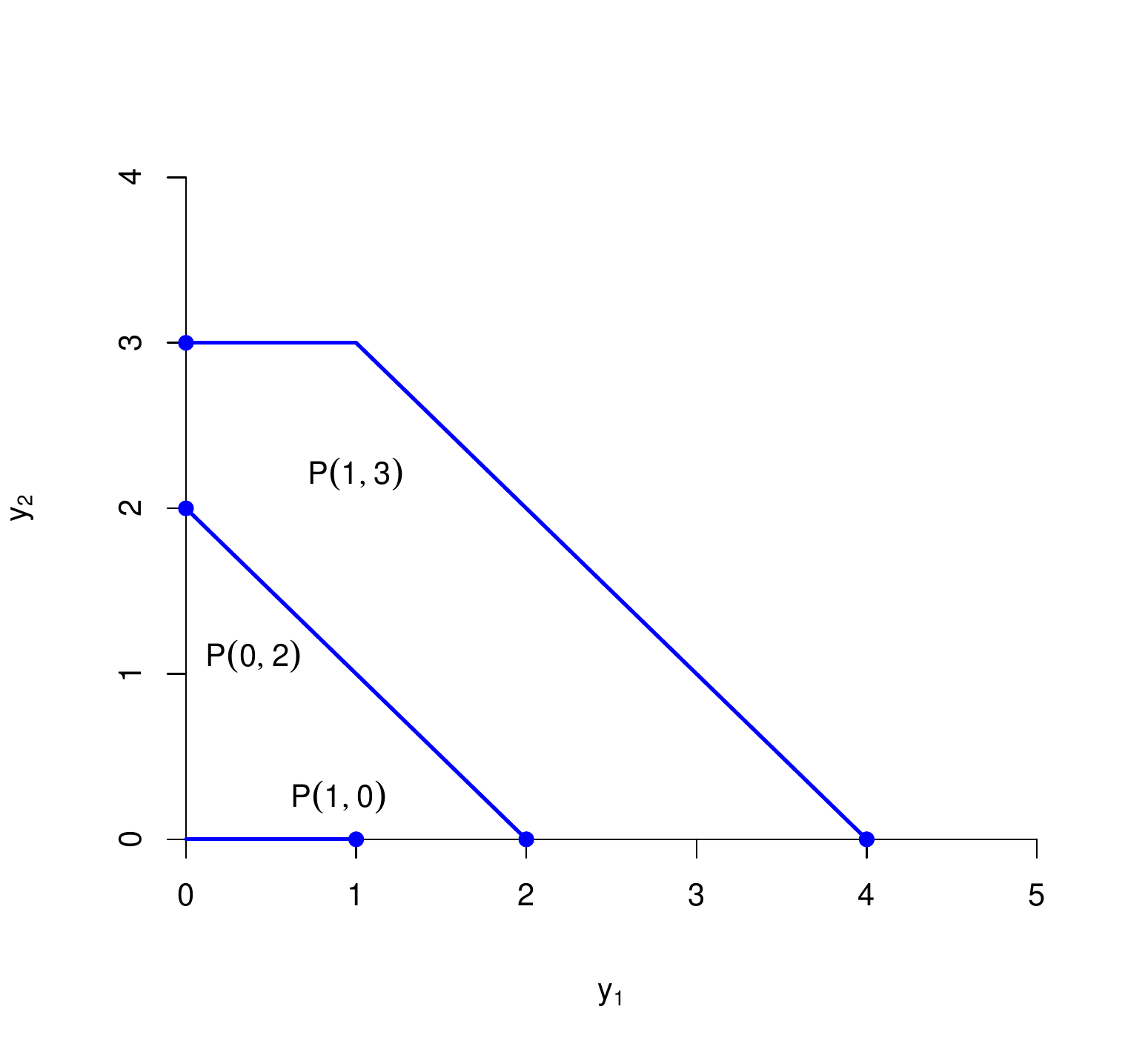}
		\caption{$P(x) = \{y \in \R^2_{+}\mid y_2 \leq x_2 \text{ and } y_1 + y_2 \leq x_1 + x_2\}$ for all $x \in \R^2_{+}$.}
	\end{center}
	\end{figure}
	\vspace{-3mm}
	Its inverse correspondence is given by $L(y) = \{x \in \R^2_{+}\mid x_2 \geq y_2 \text{ and } x_1 + x_2 \geq y_1 + y_2\}$ for all $ y \in \R^2_{+}$, and the efficient subsets of output and input sets are given by 
	\begin{equation*}
	\begin{split}
	& \text{Eff } P(x) = \{y \in \R^2_{+}\mid x_1 \leq y_1 \leq x_1 + x_2 \text{ and } y_2 = (x_1 + x_2) - y_1 \}  \text{ and } \\
	& \text{Eff } L(y) = \{x \in \R^2_{+}\mid 0 \leq x_1 \leq y_1 \text{ and } x_2 = (y_1 + y_2) - x_1\}.
	\end{split}
	\end{equation*}
	In this case, property E2 holds, since Eff $P(x)$ and Eff $P(z)$ are disjoint whenever $x \geq z$. Property E1 also holds, since Eff $L(y)$ and Eff $L(v)$ are disjoint whenever $y \geq v$. We conclude that conditions T1--T5 and T6$^\ast$--T7$^\ast$ do not imply the equality of the weak efficient and efficient subsets of an output set for each $x \in X_1$. 
	
	\textbf{Lemma 2.1.10.} Let the production possibilities set satisfy conditions T1--T5 and T6$^\ast$--T7$^\ast$. If WEff $P(x_0) \neq$ Eff $P(x_0)$ for some $x_0 \in X_1$, then $F^{i}$ is not an efficient joint production function for some $i \in \{1,\dots,m\}$.  
	
	\textsc{Proof.} Suppose there exists $x_0 \in X_1$ such that WEff $P(x_0) \neq$ Eff $P(x_0)$ and choose $y$ in WEff $P(x_0)\setminus$Eff $P(x_0)$. Since $y \in$ WEff $P(x_0)$, for all $w \in P(x_0)$, there exists $i \in \{1,\dots, m\}$ such that $w_i \leq y_i$ and $y_i > 0$. Since $y \not\in$ Eff $P(x_0)$, there exists $u \in P(x_0)$ such that $u_k > y_k$ for some $k$ and $u_j \geq y_j$ if $j \neq k$. The conjunction of two statements implies that $u_i = y_i$ and $y_i > 0$ for some $i \neq k$. Let $U = \{i \in \{1,\dots,m\}\mid u_i = y_i \text{ and } y_i > 0\}$. Then $u_j > y_j$ or $u_j = y_j = 0$ for all $j \not\in U$.
	
	Let $V_i = \{v_i \in \R_{+}\mid (y_1, \dots, v_i,\dots, y_m) \in P(x_0)\}$ and suppose that, for all $i \in U$, there exists $\tilde{v}_i \in V_i$ such that $\tilde{v}_i > y_i$. Let $\tilde{v}^{i} = (y_1, \dots, \tilde{v}_i,\dots, y_m)$ for all $i \in U$. Since $P(x)$ is convex for all $x \in \ri$, it follows that $P(x_0) \ni \theta_0 u + \sum\limits_{i \in U}\theta_i \tilde{v}^i \stackrel{\ast}{>} y$ if the weights in $\{\theta_j\mid j \in \{0\}\cup U\}$ are strictly positive and sum to 1. This contradicts $y \in$ WEff $P(x_0)$. Therefore, there exists $i \in U$ such that $v_i \leq y_i$ for all $v_i \in V_i$.  Since $y_i \in V_i$, it follows that $y_i = \max V_i = t(y^{-i}, x_0)$, implying that $F^{i}(y, x_0) = 0$, although $y \not\in$ Eff $P(x_0)$. Hence, $\{y \in \ro\mid F^{i}(y, x_0) = 0\} \not \subseteq$ Eff $P(x_0)$ for this $i$. $\qed$ 
	
	In summary, under conditions T1--T5 and T6$^\ast$--T7$^\ast$, three joint production functions exist. However, if the weak efficient and efficient subsets of an output set are not equal for some $x \in X_1$, then, for some $i$, the unsymmetric transformation function would trace the efficient subset of this output set but not exclusively. Apparently, if the equivalence of the unsymmetric transformation function and an efficient JPF is to be established, it is not possible to dispense with a stronger monotonicity condition that ensures the equality of the weak efficient and efficient subsets of an output set for each $x \in X_1$. 
	
	Färe et al.\ (1985, pp.\ 33--34) define weak efficient strict monotonicity of the input and output correspondences (T6) and prove that it is sufficient for WEff $L(y) =$ Eff $L(y)$ for all $y \in Y_1$ and for WEff $P(x) =$ Eff $P(x)$ for all $x \in X_1$, provided condition T4 holds:
	\begin{enumerate}
		\item[T6.] for all $y \in Y_1$ and for all $x \in X_1$,
			\begin{equation*}
			\begin{split}
			\text{WE1. } & \text{WEff } L(y) \cap \text{WEff } L(v) = \varnothing \ \text{ if } \ y \geq v; \\
			\text{WE2. } & \text{WEff } P(x) \cap \text{WEff } P(z) = \varnothing \ \text{ if } \ x \geq z.
			\end{split}
			\end{equation*}
	\end{enumerate} 
	\vspace{-2mm}
	Intuitively, WE1 states that if a decrease in all nonzero inputs renders some production plan infeasible, then so does an increase in any output. WE2 states that if an increase in all nonzero outputs is not feasible, then so is a decrease in any input. It is shown in Färe (1986, p.\ 674) that condition T6 implies the existence and equivalence of an isoquant, a weak efficient, and an efficient joint production functions whenever inputs and outputs are strongly disposable. 
	
	Finally, we introduce condition T7, which, together with convexity of the production possibilities set, implies that each input vector $x \in X_1$ can produce an output vector $y > 0$, that is, with all strictly positive components:
	\begin{enumerate}
		\item[T7.] for all $x \in X_1$ and for all $i \in \{1, \dots, m\}$, there exists $y \in P(x)$ such that $y_i > 0$.
	\end{enumerate}
	The output correspondence from Example 2.1.9, for instance, does not satisfy T7, since $y_2 = 0$ for all $y \in P(1,0)$. To see why this assumption is essential to the proof of the theorem that follows, suppose there exists $i \in \{1,\dots, m\}$ such that $y_i = 0 \text{ for all } y \in P(x_0)$ for some $x_0 \in X_1$. Choose $y$ in $P(x_0)\setminus$Eff $P(x_0)$, for example, $(1/2, 0)$ in $P(1,0)\setminus\{(1,0)\}$. Then $y_i = 0$ and $t(y^{-i}, x_0) = \max\{0\} = 0$, implying that $F^{i}(y, x_0) = 0$. Therefore, the set of output--input vectors upon which $F^i$ assumes the value zero may contain not only the set upon which an efficient JPF assumes the value zero but also feasible bundles $(y,x)$ such that $y \not\in$ Eff $P(x)$ if maximization is performed with respect to an output that cannot be produced with a given input vector $x \in X_1$. 
	
	Before we prove our main result, it is of interest to determine the types of production processes that are compatible with conditions T1--T7 imposed on the production possibilities set. Condition T7 entails that, for a particular input vector, either none of the outputs can be produced (if $x \not\in X_1$) or all of them can (if $x \in X_1$). Frisch (1965, pp.\ 10--11) discusses two types of production for which this requirement is fulfilled: joint and assorted. Joint production arises when ``outputs are \textit{necessarily} produced together for physical, chemical or technical reasons'' (Baumgärtner, 2000, p.\ 7), and zero quantity of some output, along with positive quantities of the other outputs, is feasible only when the produced amount of an unwanted output is disposed of at no cost; see also Lloyd (1983, p.\ 46). When production is joint, the output mix can be either fixed or varied to a certain extent for a given input vector (Frisch, 1965, p.\ 11). In other words, the efficient subset of an output set is not necessarily a singleton; however, it does not contain any output vectors with one or more zero components. This implies that, for all $x \in X_1$, the weak efficient and efficient subsets of an output set are not equal, and therefore condition T6 does not hold. It follows from the previous discussion that the unsymmetric transformation function and any of the joint production functions are not equivalent if outputs are produced jointly.
	
	Assorted production arises, according to Frisch (1965), when, for a given input vector, there is a choice regarding which output to produce. In the two-output case, all resources, for example, can be allocated to the production of only the first output or only the second or some combination of the two (pp.\ 276--277). Hence, there is maximal flexibility in the choice of relative output quantities (Lynne, 1974, p.\ 55); that is, a given input vector can be used to produce any output mix efficiently. This type of production is compatible with conditions T1--T7. Next, we prove their sufficiency for the equivalence of the unsymmetric transformation function and an efficient JPF.
	
	\textbf{Theorem 2.1.11.} If the production possibilities set satisfies conditions T1--T7, then the unsymmetric transformation and an efficient joint production functions are equivalent. 
	
	\textsc{Proof.} Since conditions T4 and T6 imply the existence and equivalence of three joint production functions, it suffices to show that
	\begin{equation*}
 	\{(y,x)\in (\ro\times X_1)\cup(Y_1\times\ri)\mid F^{i}(y,x) = 0\} = \{(y,x)\in Y_1\times X_1\mid y \in \text{Eff } P(x)\}
	\end{equation*}
	for all $i \in \{1,\dots,m\}$; see (2.1.5) and (2.1.3). First, note that from the definition of $t$ it follows that $F^{i}(y,x) \neq 0$ for all $(y,x) \in (Y_2\times X_1)\cup(Y_1\times X_2)\cup(Y_1\times X_3)$ and for all $i \in \{1,\dots,m\}$, whereas conditions T7 and T4 imply that this result remains valid for all $(y,x) \in (Y_3\times X_1)$. Hence, for all $i \in \{1,\dots,m\}$,
	\begin{equation*}
	\{(y,x)\in ((Y_2\cup Y_3)\times X_1)\cup(Y_1\times (X_2\cup X_3))\mid F^{i}(y,x) = 0\} = \varnothing.
	\end{equation*}
	Let $(y,x) \in Y_1\times X_1$, choose $i$ from $\{1,\dots,m\}$, and suppose $F^{i}(y, x) = 0$. Then $y_i = t(y^{-i}, x)$ and either $y_i > 0$ or $y_i = 0$.
	
	\textbf{Case 1.} Suppose $y_i = t(y^{-i}, x)$ and $y_i > 0$. From the definition of $t$ it follows that $(y_1,\dots, v_i,\dots, y_m) \not\in P(x)$ if $v_i > y_i$, which together with strong disposability of outputs implies that $v \not\in P(x)$ if $v \stackrel{\ast}{>} y$. Consequently, $y \in$ Eff $P(x)$, since Eff $P(x)$ = WEff $P(x)$ whenever conditions T4 and T6 hold.
	
	\textbf{Case 2.} Suppose $y_i = t(y^{-i}, x)$, $y_i = 0$, and $y \not\in$ Eff $P(x)$. By condition T7, there exists $u \in P(x)$ such that \mbox{$u_i > 0$}. Since $y \not\in$ WEff $P(x)$, there also exists $w \in P(x)$ such that $w \stackrel{\ast}{>} y$. Convexity of $P(x)$ implies that $\tau w + (1 - \tau)u \in P(x)$ for all $\tau \in [0,1]$. Let $\widetilde\tau = \max \{\tau_j\mid \tau_j = y_j/w_j \text{ if } j \neq i \text{ and } w_j \neq 0\}$. Since $y \geq 0$ and $w \stackrel{\ast}{>} y$, it follows that $\widetilde\tau \in (0, 1)$. Furthermore, $\widetilde\tau w + (1 - \widetilde\tau)u \geq y$ and $\widetilde u_i = \widetilde\tau w_i  + (1 - \widetilde\tau)u_i > 0$. By strong disposability of outputs, $(y_1,\dots,\widetilde u_i,\dots, y_m) \in P(x)$ and, hence, $t(y^{-i}, x) \geq \widetilde u_i > 0 = y_i$, which leads to a contradiction.  
	
	Conversely, suppose $F^{i}(y,x) \neq 0$ for some $i \in \{1,\dots,m\}$. If \mbox{$y_i > t(y^{-i}, x)$}, then $y \not\in P(x)$, which includes Eff $P(x)$. If $y_i < t(y^{-i}, x)$, then $P(x) \ni (y_1, \dots, t(y^{-i}, x),\dots, y_m) \geq y$, implying that $y \not\in $ Eff $P(x)$. Therefore, if $y \in$ Eff $P(x)$, then $F^{i}(y,x) = 0$ for each $i \in \{1, \dots, m\}$. $\qed$
	
	The properties of the unsymmetric transformation function that hold under conditions T1--T5 can be found in Diewert (1973, p.\ 287). If $T$ satisfies, in addition, T6 and T7, then monotonicity properties of $t$ can be strengthened to $t$ being strictly increasing in inputs, i.e., if $z\geq x$ and $t(y^{-i},z) > -\infty$, then $t(y^{-i},z) > t(y^{-i},x)$, and strictly decreasing in outputs, i.e., if $v^{-i} \leq y^{-i}$ and $t(v^{-i},x) > -\infty$, then $t(v^{-i},x) > t(y^{-i},x)$, for an arbitrary $i$. In the next section, we extend the discussion to a symmetric transformation function and exhibit the properties of this function that are sufficient for the production possibilities set induced by it to satisfy conditions T1--T7. We also prove that a symmetric transformation function is an efficient joint production function under these assumptions. 

	\subsection{Symmetric transformation function} 
	
	Let a symmetric transformation function $F\colon \roi \to \R$ represent production technology and assume that $F$ satisfies properties F1--F4, which partly overlap with the ones suggested by Hanoch (1970, p.\ 423) and Lau (1972, p.\ 281):
	\begin{enumerate}
		\item[F1.] $F(0,0) = 0$;
		\item[F2.] $F$ is a continuous function;
		\item[F3.] $F$ is strictly decreasing in inputs, i.e., $F(y, x') < F(y,x)$ if $x' \geq x$, and strictly increasing in outputs, i.e., $F(y',x) > F(y,x)$ if $y' \geq y$;
		\item[F4.] $F$ is a convex function.
	\end{enumerate}
	\textbf{Theorem 2.2.1.} If a transformation function $F$ satisfies properties F1--F4, then the set 
	\begin{equation}\tag{2.2.2}
	\, T = \{(y,x) \in \roi\mid F(y,x) \leq 0\}
	\end{equation}
	satisfies properties T1--T7:
	\begin{enumerate}
		\item[T1.] $T$ is a nonempty subset of $\roi$; in particular, $(0,0) \in T$;
		\item[T2.] $T$ is closed;
		\item[T3.] $T$ is convex;
		\item[T4.] if $(y, x) \in T$ and $(-y',x') \geqq (-y,x)$, then $(y',x') \in T$;
		\item[T5.] $P(x)$ is bounded for all $x \in X_1$;
		\item[T6.] for all $y \in Y_1$ and for all $x \in X_1$,
			\begin{equation*}
			\begin{split}
			\text{WE1. } & \text{WEff } L(y) \cap \text{WEff } L(v) = \varnothing \text{ if } y \geq v; \\
			\text{WE2. } & \text{WEff } P(x) \cap \text{WEff } P(z) = \varnothing \text{ if } x \geq z;
			\end{split}
			\end{equation*}
	\end{enumerate}
	\begin{enumerate}
		\item[T7.] for all $x \in X_1$ and for all $i \in \{1, \dots, m\}$, there exists $y \in P(x)$ such that $y_i > 0$.
	\end{enumerate}
	\textsc{Proof.} T1, T2, T3, and T4 follow from F1, F2, F4, and F3, respectively, and the definition of $T$, whereas property T7 follows from F1--F3.
	
	Next, we prove that if $F$ satisfies properties F1--F4, then $T$ satisfies property T5. Let $x \in X_1$ and define $P(x) = \{y \in \ro\mid F(y,x) \leq 0\}$. Let $e_i$ denote the vector in $\ro$ that has the $i$th component equal to 1 and the other components equal to 0. For each $i \in \{1,\dots,m\}$, consider the function $g_i\colon \R_{+} \to \R$ given by $g_i(\mu) = F(\mu e_i, x)$, which is the restriction of $F$ to the ray $\{(0,x) + \mu(e_i,0)\mid \mu \in \R_{+}\}$. Properties F1--F4 imply that $g_i$ is a continuous, strictly increasing, and convex function with $g_i(0) < 0$. Therefore, for all $i\in \{1,\dots, m\}$, there exists $\mu_i^\ast > 0$ such that $g_i(\mu^\ast_i) = 0$. From this result and property F3, it follows that if $F(y,x) \leq 0$, then $y \in C = \{y \in \ro\mid 0 \leq y_i \leq \mu_i^\ast \text{ for each } i \in \{1,\dots,m\}\}$. Since the set $P(x)$ is included in the closed cell $C$, we conclude that $P(x)$ is bounded. 
	
	We proceed to show that $T$ satisfies property T6 if $F$ satisfies properties F2 and F3. Let $L(y) = \{x \in \ri\mid F(y,x) \leq 0\}$ and suppose that $y \geq v$ and $x \in \text{WEff } L(y)$. Set $\varepsilon = |F(v,x)/2|$, where $F(v,x) < F(y,x) \leq 0$ by property F3. Since $F$ is continuous, there exists $\delta > 0$ such that $F(B_\delta(v,x)\cap (\roi)) \subseteq B_\varepsilon (F(v,x))$. Let $\lambda^\ast = \left(1 - \delta/(2\norm{x})\right)$ if $\delta/(2\norm{x}) < 1$, and $\lambda^\ast = 1/2$ otherwise. It follows that $\lambda^\ast \in (0,1)$, $\lambda^\ast x \stackrel{\ast}{<} x$, and $(v, \lambda^\ast x) \in B_\delta(v,x)\cap (\roi)$, which implies that $F(v, \lambda^\ast x) < 0$. Thus, $\lambda^\ast x \in L(v)$ and, consequently, $x \not\in \text{WEff } L(v)$. A similar argument shows that properties F2 and F3 imply WE2.  $\qed$ 
	
	As discussed in Section 2.1, if the production possibilities set satisfies properties T1--T7, then three joint production functions exist and the subsets of $(\ro\times X_1)\cup(Y_1\times\ri)$ upon which they assume the value zero are equal. Our final goal is to show that a symmetric transformation function is an efficient joint production function whenever properties F1--F4 hold.  
	
	\textbf{Theorem 2.2.3.} If a symmetric transformation function $F$ satisfies properties F1--F4 and the production possibilities set is defined by (2.2.2), then $F$ is an efficient joint production function.
	
	\textsc{Proof.} First, note that $X_2 = \varnothing$ if properties F1--F3 hold. As in the proof of Theorem 2.1.11, we need to show that 
	\begin{equation*}
	\{(y,x)\in (\ro\times X_1)\cup(Y_1\times X_3)\mid F(y,x) = 0\} = \{(y,x)\in Y_1\times X_1\mid x \in \text{Eff } L(y)\}.
	\end{equation*}
	Properties F1 and F3 and the definition of $T$ imply that 
	\begin{equation*}
	\{(y,x)\in (Y_2\times X_1)\cup(Y_3\times X_1)\cup(Y_1\times X_3)\mid F(y,x) = 0\} = \varnothing.
	\end{equation*}
	Let $(y,x) \in Y_1\times X_1$ and suppose $F(y,x) = 0$. Since $x \in L(y)$ and, by property F3, $F(y,x') > 0$ if $x' \leq x$, it follows that $x \in$ Eff $L(y)$. 
	
	To prove the converse, we partly follow Bol and Moeschlin (1975, p.\ 398). Suppose that $x \in$ Eff $L(y)$ and $F(y,x) < 0$. Consider a sequence $\{\lambda_n\}$ in $(0,1)$ that converges to~1. Then~$\lambda_n x \stackrel{\ast}{<} x$ for all $n \in \mathbb{N}$ and the sequence $\{F(y,\lambda_n x)\}$ converges to $F(y,x)$ by continuity of $F$. It follows that $F(y,\lambda_n x) < 0$ for all but finitely many $n \in \mathbb{N}$, which contradicts $x \in$ Eff $L(y)$. Therefore, $F(y,x) = 0$ whenever $x \in$ Eff $L(y)$. $\qed$
	
	Lastly, we note that the same argument can be used to verify the assertion of Hanoch (1970, p.\ 423) that the efficient subset of $T$ is characterized by the equation $F(y,x) = 0$. In particular, let Eff $T = \{(y,x) \in T \mid (y', x') \not\in T \text{ if } (-y', x') \leq (-y, x)\}$.  Färe et~al.\ (1985, p.\ 47)  show that conditions T4 and T7$^\ast$ from Section 2.1 imply that $(y,x) \in$ Eff~$T$ if and only if $x \in$ Eff $L(y)$ and $y \in$ Eff $P(x)$ for all $(y,x) \in Y_1\times X_1$. This result, however, remains true when monotonicity condition T7$^\ast$ is dropped. It then follows from Theorem~2.2.3 that $F(y,x) = 0$ if and only if $(y,x) \in \text{Eff } T$ for all $(y,x) \in \roi$ under conditions F1--F4.  

	\section{Directional technology distance function}
	\vspace{-2mm}
	\subsection{Quadratic function}
	
	Let $(g_y, g_x)$ denote a nonzero direction vector, where $g_y \in \ro$ and $g_x \in \ri$. Adapting the benefit and shortage functions of Luenberger (1992a, 1992b) to the production and efficiency measurement contexts, Chambers (1996) and Chambers et al.\ (1996, 1998) provide the following definition of the directional technology distance function $\oset{D}_T\colon \rp\times(\rp\setminus\{(0,0)\}) \to \R\cup\{-\infty\}$. 
	
	\textbf{Definition 3.1.1 (Chambers, Chung, and Färe, 1998).} For all $(y,x) \in \rp$ and for all $(g_y, g_x) \in \rp\setminus\{(0,0)\}$, 
	\begin{equation*}
	\oset{D}_T(y,x; g_y, g_x) = \begin{cases} \sup\{\beta \in \R\mid (y + \beta g_y,x - \beta g_x) \in T\} & $if $ (y + \beta g_y,x - \beta g_x) \in T \\ & $for some $ \beta \in \R; \\ -\infty & $otherwise$. \end{cases}
	\end{equation*} 
	Chambers (1996) and Chambers et al.\ (1996, 1998), among others, establish two properties of the directional technology distance function that follow directly from its definition: translation property (D1) and homogeneity of degree $-1$ in the direction vector (D2). This result is summarized in Proposition 3.1.2.
	
	\textbf{Proposition 3.1.2.} If $\oset{D}_T$ is the directional technology distance function, then $\oset{D}_T$ satisfies properties D1 and D2 for all $(y,x) \in \rp$ and for all $(g_y, g_x) \in \rp\setminus\{(0,0)\}$:
	\begin{enumerate}
		\item[D1.] $\oset{D}_T(y + \alpha g_y,x - \alpha g_x; g_y, g_x) = \oset{D}_T(y,x; g_y, g_x) - \alpha$ for every $\alpha \in \R$ that satisfies $(y + \alpha g_y,x - \alpha g_x) \in \rp$;
		\item[D2.] $\oset{D}_T(y,x; \psi g_y, \psi g_x) = \psi^{-1}\oset{D}_T(y,x; g_y, g_x)$ for all $\psi > 0$.
	\end{enumerate}
	\textsc{Proof.} See Luenberger (1992a, p.\ 464), Chambers et al.\ (1996, p.\ 416), and Hudgins and Primont (2007, p.\ 40). $\qed$ 
	
	Equivalently, if a function does not satisfy property D1 \textbf{or} property D2, then it is not the directional technology distance function. Next, we show that a quadratic function that is restricted to satisfy translation property is not homogeneous of degree $-1$ in the direction vector, and therefore it is not the DTDF by Proposition 3.1.2. Since this is the only (to the best of our knowledge) functional form that is used in econometric estimation of the systems of simultaneous equations including the DTDF, this necessitates the search for alternative functional forms that satisfy both properties D1 and D2.
	
	Chambers (1996, pp.\ 14--18) and Hudgins and Primont (2007, pp.\ 38--41) consider a quadratic function of inputs and outputs 
	
	{\centering
	$ \displaystyle
	Q(y,x) = \alpha_0 + \sum_{i=1}^{n}\alpha_i x_i + \sum_{k=1}^{m}\beta_k y_k + \frac{1}{2}\sum_{i=1}^{n}\sum_{j=1}^{n}\alpha_{ij}x_i x_j + \frac{1}{2}\sum_{k=1}^{m}\sum_{\ell=1}^{m}\beta_{k\ell}y_k y_\ell + \sum_{i=1}^{n}\sum_{k=1}^{m}\gamma_{ik}x_i y_k
	$ 
	\par} 

	and impose restrictions on its parameters that incorporate the direction vector and ensure that the resulting function of inputs, outputs, and directions $Q(y,x; g_y,g_x)$ satisfies translation property D1. In particular, in addition to the symmetry restrictions $\alpha_{ij} = \alpha_{ji}$ for all $i,j \in \{1, \dots, n\}$ and $\beta_{k\ell} = \beta_{\ell k}$ for all $k,\ell \in \{1, \dots, m\}$, Chambers (1996, p.\ 14) imposes the following translation restrictions: for all $i \in \{1, \dots, n\}$,
	\begin{equation*}
	\sum_{k=1}^{m}\gamma_{ik}g_{yk} - \sum_{j=1}^{n}\alpha_{ij}g_{xj} = 0; \quad \sum_{k=1}^{m}\beta_k g_{yk} - \sum_{i=1}^{n}\alpha_i g_{xi} = -1; \, \text{  and  } \, \sum_{\ell=1}^{m}\beta_{k\ell}g_{y\ell} - \sum_{i=1}^{n}\gamma_{ik}g_{xi} = 0
	\end{equation*}
	for all $k \in \{1, \dots, m\}$.
	
	We follow Atkinson and Tsionas (2016, p.\ 303) in solving these restrictions for the parameters
	\begin{equation*}
	\begin{split}
	& \alpha_n =  \frac{1}{g_{xn}}\left(\sum_{k=1}^{m}\beta_k g_{yk} - \sum_{i=1}^{n-1}\alpha_i g_{xi}  + 1 \right), \\
	& \alpha_{in} = \alpha_{ni} = \frac{1}{g_{xn}}\left(\sum_{k=1}^{m}\gamma_{ik} g_{yk} - \sum_{j=1}^{n-1}\alpha_{ij} g_{xj}\right) \text{ for all } i \in \{1, \dots, n-1\}, \\
	& \alpha_{nn} = \frac{1}{g_{xn}}\left(\sum_{k=1}^{m}\gamma_{nk}g_{yk} - \sum_{j=1}^{n-1}\left(\sum_{k=1}^{m}\gamma_{jk}g_{yk} - \sum_{p=1}^{n-1}\alpha_{jp}g_{xp}\right)\frac{g_{xj}}{g_{xn}}\right), \\
	& \beta_{km} = \beta_{mk} = \frac{1}{g_{ym}}\left(\sum_{i=1}^{n}\gamma_{ik}g_{xi} - \sum_{\ell=1}^{m-1}\beta_{k\ell}g_{y\ell}\right)  \text{ for all } k \in \{1, \dots, m-1\}, \\
	& \beta_{mm} = \frac{1}{g_{ym}}\left(\sum_{i=1}^{n}\gamma_{im}g_{xi} - \sum_{\ell=1}^{m-1}\left(\sum_{i=1}^{n}\gamma_{i\ell}g_{xi} - \sum_{r=1}^{m-1}\beta_{\ell r}g_{yr}\right)\frac{g_{y\ell}}{g_{ym}}\right) 
	\end{split}
	\end{equation*}
	
	and in incorporating them into the quadratic function of inputs and outputs to obtain 
	\begin{equation*}
	\begin{split}
	Q(y,x; & \ g_y,g_x) =  \alpha_0 + \sum_{i=1}^{n-1}\alpha_i \left(x_i - \frac{g_{xi}}{g_{xn}}x_n\right) + \sum_{k=1}^{m}\beta_k \left(y_k + \frac{g_{yk}}{g_{xn}}x_n\right) + \frac{x_n}{g_{xn}} \\
	& + \sum_{i=1}^{n-1}\sum_{j=i}^{n-1}\left[\frac{1}{2}\right]^{\mathbbm{1}(i = j)}\alpha_{ij}\left(x_i - \frac{g_{xi}}{g_{xn}} x_n\right)\left(x_j - \frac{g_{xj}}{g_{xn}} x_n\right)  \\
	& + \sum_{k=1}^{m-1}\sum_{\ell=k}^{m-1}\left[\frac{1}{2}\right]^{\mathbbm{1}(k = \ell)}\beta_{k\ell}\left(y_k - \frac{g_{yk}}{g_{ym}} y_m\right)\left(y_\ell - \frac{g_{y\ell}}{g_{ym}} y_m\right) \\
	& + \sum_{i=1}^{n}\sum_{k=1}^{m}\gamma_{ik}\left(x_i + \frac{g_{xi}}{g_{ym}}y_m\right)\left(y_k + \frac{g_{yk}}{g_{xn}}x_n\right) - \frac{1}{2}\sum_{i=1}^{n}\sum_{k=1}^{m}\gamma_{ik} g_{xi} g_{yk}\left(\frac{x_n}{g_{xn}} + \frac{y_m}{g_{ym}}\right)^2.
	\end{split}
	\end{equation*}

	This function is not homogeneous of degree $-1$ in the direction vector and, consequently, is not the directional technology distance function. Since functional forms satisfying both properties D1 and D2 are not readily available, we suggest in Section~3.2 an alternative approach whereby the directional distance function is derived from a symmetric transformation function, discussed in Section 2.2.

	\subsection{DTDF derived from symmetric transformation function}
	
	Consider the following parametric optimization problem:
	\begin{equation}\tag{P}
	\begin{split}
	\begin{cases}
	& \quad \text{Maximize } \beta \qquad \text{subject to } \\
	& \, -\beta g_{yj} - y_j \leq 0 \quad \text{ for all } \ j \in \{1, \dots, m\}, \\
	& \quad \beta g_{xi} - x_i \leq 0 \quad \text{ for all } \ i \in \{1, \dots, n\}, \\
	& \quad F(y + \beta g_y, x - \beta g_x) \leq 0, \\
	& \quad \beta \in \R, \end{cases}
	\end{split}
	\end{equation} 
	where $(y,x) \in \rp$ and $(g_y, g_x) \in \rp\setminus\{(0,0)\}$ are parameters and $F$ is a symmetric transformation function satisfying properties F1--F4. It follows from Definition 3.1.1 that, for each $(y,x) \in \rp$ and $(g_y, g_x) \in \rp\setminus\{(0,0)\}$, the optimal value of (P) is equal to the value of the directional technology distance function associated with the production possibilities set $T$ defined by (2.2.2). 
	
	Our goal now is to find the optimal value of the problem (P) for each $(y,x) \in \rp$ and $(g_y, g_x) \in \rp\setminus\{(0,0)\}$. To this end, we next define correspondences $\mathrm{I}^{+}$, $\mathrm{J}^{+}$, $\Gamma$, $S$, and $\Lambda$, but to simplify notation we drop the dependence of their image sets on $(y,x)$ and/or $(g_y, g_x)$ whenever these vectors are fixed throughout the discussion. 
	
	Fix $(y,x) \in \rp$ and $(g_y, g_x) \in \rp\setminus\{(0,0)\}$ and let $\mathrm{I}^{+} =\{i \in \{1,\dots, n\}\mid g_{xi} > 0\}$ and $\mathrm{J}^{+} = \{j \in \{1,\dots,m\}\mid g_{yj} > 0\}$. When we discard the redundant constraints, the optimal value of (P) becomes
	\begin{equation}\tag{3.2.1}
	\sup\{\beta \in [\sup\limits_{j \in \mathrm{J}^{+}}\{-y_j/g_{yj}\}, \inf\limits_{i \in \mathrm{I}^{+}}\{x_i/g_{xi}\}]\setminus\{-\infty, +\infty\} \mid F(y + \beta g_y,x - \beta g_x) \leq 0\}.
	\end{equation}
	Here, we follow the convention that $\sup\varnothing = -\infty$ and $\inf\varnothing = +\infty$. 
	
	Let $\Gamma$ denote the set $[\sup\limits_{j \in \mathrm{J}^{+}}\{-y_j/g_{yj}\}, \inf\limits_{i \in \mathrm{I}^{+}}\{x_i/g_{xi}\}]\setminus\{-\infty, +\infty\}$ and consider the function $F_{[S]}\colon \Gamma \to \R$ given by $F_{[S]}(\beta) = F(y + \beta g_y,x - \beta g_x)$ for all $\beta \in \Gamma$. The function $F_{[S]}$ is the restriction of $F$ to the line segment or the ray $S = \{(y,x) + \beta (g_y,-g_x)\mid \beta \in \Gamma\}$, included in $\rp$. Properties F2--F4 imply that $F_{[S]}$ is a continuous, strictly increasing, and convex function, and therefore it has a left inverse. That is, there exists a function $G_S\colon \R \to \Gamma$ such that $G_S \circ F_{[S]} = \mathrm{Id}_{\Gamma}$, where $\mathrm{Id}_{\Gamma}\colon \Gamma \to \Gamma$ is the identity function on $\Gamma$. 
	
	Also, let $\Lambda = \{\beta \in \Gamma\mid F_{[S]}(\beta) \leq 0\}$. There are three cases to consider. First, if $\Lambda = \varnothing$, then $F_{[S]}(\beta) > 0$ for all $\beta \in \Gamma$ and no output--input bundle on the ray or the line segment passing through the point $(y,x)$ in the direction $(g_y, -g_x)$ is feasible. Otherwise, $\Lambda \neq \varnothing$ and either $\Lambda = \Gamma$ or $\Lambda \subset \Gamma$. It is shown next that if the sets $\Lambda$ and $\Gamma$ are equal, then the set $\Gamma$ must be bounded above, whereas if $\Lambda$ is a proper subset of $\Gamma$, then $\sup\Lambda = G_S(0)$. 
	\newpage
	
	\textbf{Lemma 3.2.2.} Let a transformation function $F$ satisfy properties F1--F4. If $\Lambda = \Gamma$, then $\sup\Lambda = \min\limits_{i \in \mathrm{I}^{+}}\left\{x_{i}/g_{xi}\right\}$. 
	
	\textsc{Proof.} Suppose $F_{[S]}(\beta) \leq 0$ for all $\beta \in \Gamma$ and $g_x = 0$. It follows that $F(y + \beta g_y, x) \leq 0$ for all $\beta \in [\max\limits_{j \in \mathrm{J}^{+}}\{-y_j/g_{yj}\}, \infty)$, which implies that $P(x)$ is not bounded. This, however, contradicts condition T5. Therefore, $g_x \neq 0$ whenever $\Lambda = \Gamma$, and consequently $\sup \Lambda = \min\limits_{i \in \mathrm{I}^{+}}\left\{x_{i}/g_{xi}\right\}$. $\qed$
	
	\textbf{Lemma 3.2.3.} Let a transformation function $F$ satisfy properties F1--F4. If $\Lambda \neq \varnothing$ and $\Lambda \subset \Gamma$, then $\sup\Lambda = G_S(0)$.
	
	\textsc{Proof.} Suppose that $\Lambda$ is a nonempty proper subset of $\Gamma$. Then there exist $\beta_1, \beta_2 \in \Gamma$ such that $F_{[S]}(\beta_1) \leq 0$ and $F_{[S]}(\beta_2) > 0$. Connectedness of $\Gamma$ and continuity of $F_{[S]}$ imply that the range of $F_{[S]}$ is connected, and therefore $0 \in$ ran $F_{[S]}$; that is, there exists $\beta^\ast \in \Gamma$ such that $F_{[S]}(\beta^\ast) = 0$. Also, suppose that $\sup \Lambda \neq G_S(0)$. Since $G_S(0) = G_S(F_{[S]}(\beta^\ast)) =$ $[G_S \circ F_{[S]}](\beta^\ast) = \mathrm{Id}_{\Gamma}(\beta^\ast) = \beta^\ast$, it follows that $\beta^\ast < \sup\Lambda$. Hence, there exists $\tilde{\beta} \in \Lambda$ such that $\tilde{\beta} > \beta^\ast$, which contradicts $F_{[S]}$ being strictly increasing.  Therefore, $\sup\Lambda = G_S(0)$ whenever $\Lambda \neq \varnothing$ and $\Lambda \subset \Gamma$. $\qed$
	
	We summarize these results in Theorem 3.2.4.
	
	\textbf{Theorem 3.2.4.} Let a transformation function $F$ satisfy properties F1--F4. Fix $(y,x) \in \rp$ and $(g_y, g_x) \in \rp\setminus\{(0,0)\}$ and let $\oset{D}_{\kern-1pt F}(y,x; g_y, g_x)$ denote the optimal value of the parametric optimization problem (P). Then
	\begin{equation}\tag{3.2.5}
	\begin{split}
	\oset{D}_{\kern-1pt F}(y,x; g_y, g_x) = \begin{cases} G_S(0) & $if $\Lambda \neq \varnothing $ and $ \Lambda \subset \Gamma; \\
	\min\limits_{i \in \mathrm{I}^{+}}\left\{x_{i}/g_{xi}\right\} & $if $\Lambda \neq \varnothing $ and $ \Lambda = \Gamma; \\
	-\infty & $if $\Lambda = \varnothing.\end{cases}
	\end{split}
	\end{equation}

	\textsc{Proof.} Theorem 3.2.4 follows from Lemmas 3.2.2 and 3.2.3 and from the discussion above. $\qed$ 
	
	To gain some intuition behind the preceding results, consider the directional technology distance function $\oset{D}_{\kern-1pt F}$ given by (3.2.5), where the subscript indicates its dependence on a symmetric transformation function $F$. As discussed by Chambers et al.\ (1998, p.\ 354), for a feasible output--input bundle $(y,x)$, the DTDF returns the distance from $(y,x)$ to its projection onto the boundary of the production possibilities set $T$ defined by (2.2.2) in a direction $(g_y, -g_x)$, if the norm of the direction vector equals unity. However, the projection of $(y,x)$ onto the boundary of $T$ may or may not be in the efficient subset of $T$. If the projection belongs to Eff $T$, then the transformation function constraint $F(y + \beta g_y, x - \beta g_x) \leq 0$ in (P) is saturated at the optimum $\beta^\ast$, i.e., $F(y + \beta^\ast g_y, x - \beta^\ast g_x) = 0$. This follows from Theorem 2.2.3 and Lemma 3.2.3. If the projection does not belong to Eff $T$, then the ray or the line segment $S$ does not intersect Eff $T$ and $(y,x)$ is projected onto a subset of the boundary of $\rp$ that contains input vectors with one or more (but not~all) zero components. Figure~3 illustrates these two cases when $g_y = 0$. Also, note that $\oset{D}_{\kern-1pt F}(y,x; g_y, g_x) = 0$ does not imply that $(y,x) \in \text{Eff } T$. 
	
	\bigskip
	\begin{figure}[H]  
	\begin{center}
		\includegraphics[width = 0.7\textwidth,trim={1.6cm 2cm 1cm 3.2cm},clip]{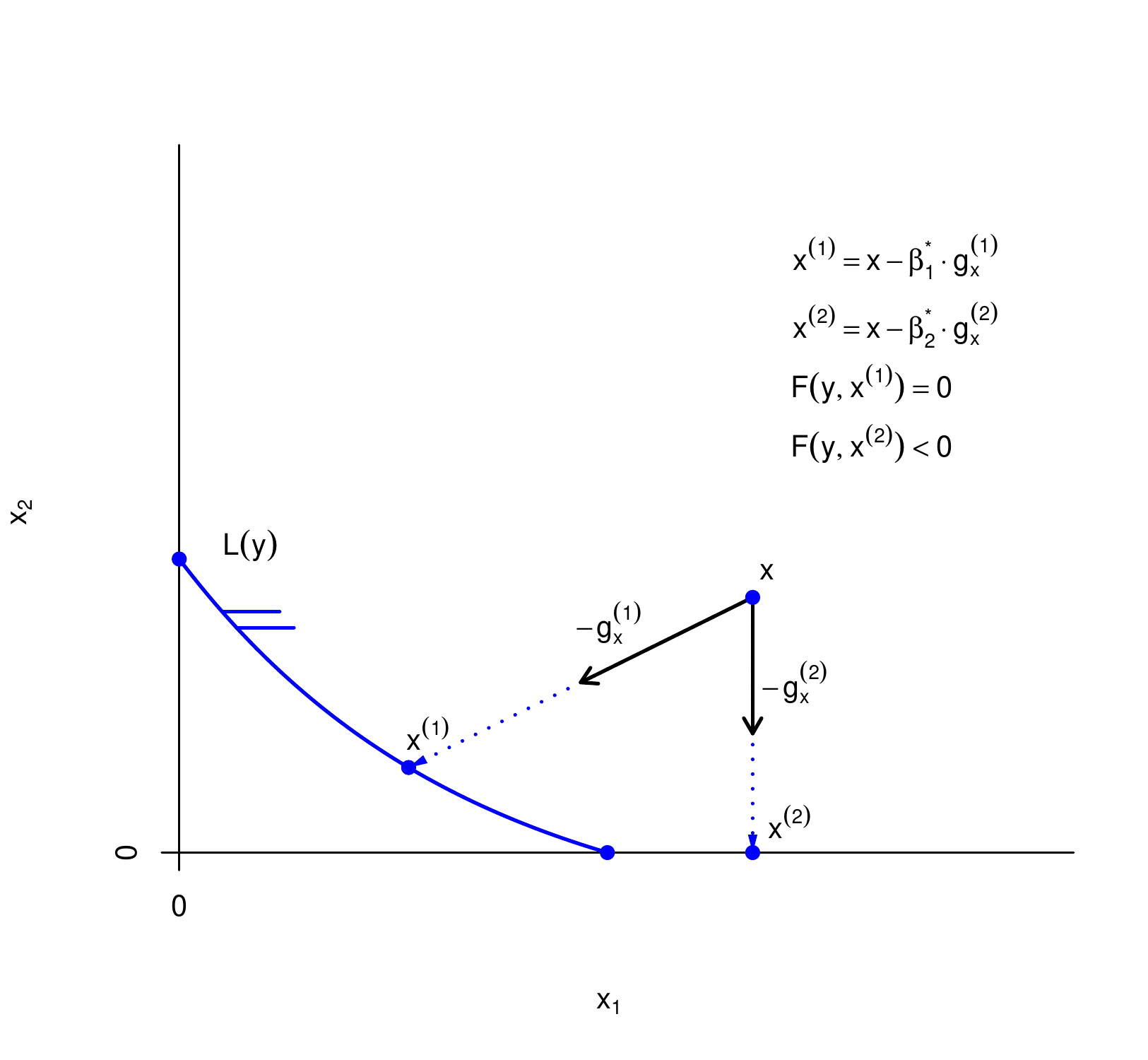}
		\caption{The projection of $(y,x)$ onto the boundary of $T$  in the direction $(0,-g_x^{(2)})$ does not belong to Eff $T$. In this case, $\beta_2^\ast = \protect\oset{D}_{\kern-1ptF}(y,x; 0, g_x^{(2)}) = \min\big\{ x_{i}/g^{(2)}_{xi}\mid i \in \mathrm{I}^{+}\big(g_x^{(2)}\big)\big\}$.}
	\end{center}
	\end{figure}
	\vspace{-3mm}
	Some of the properties of $\oset{D}_{\kern-1ptF}$, akin to those stated by Chambers (1996) and Chambers et al.\ (1996, 1998), are summarized in Theorem 3.2.6.
	
	\textbf{Theorem 3.2.6.} Let a transformation function $F$ satisfy properties F1--F4 and the directional technology distance function $\oset{D}_{\kern-1ptF}$ be given by (3.2.5). Then $\oset{D}_{\kern-1ptF}$ satisfies properties D1--D6 for all $(y,x) \in \rp$ and for all $(g_y, g_x) \in \rp\setminus\{(0,0)\}$:
	\begin{enumerate}
		\item[D1.] $\oset{D}_{\kern-1ptF}(y + \alpha g_y,x - \alpha g_x; g_y, g_x) = \oset{D}_{\kern-1ptF}(y,x; g_y, g_x) - \alpha$ for every $\alpha \in \R$ that satisfies $(y + \alpha g_y,x - \alpha g_x) \in \rp$;
		\item[D2.] $\oset{D}_{\kern-1ptF}(y,x; \psi g_y, \psi g_x) = \psi^{-1}\oset{D}_{\kern-1ptF}(y,x; g_y, g_x)$ for all $\psi > 0$;
		\item[D3.] $\oset{D}_{\kern-1ptF}(0,0; g_y, g_x) = 0$; 
		\item[D4.] $\oset{D}_{\kern-1ptF}(y,x; g_y, g_x) \geq 0$ if and only if $F(y,x) \leq 0$;
		\item[D5.] $\oset{D}_{\kern-1ptF}$ is nondecreasing in inputs, i.e., $\oset{D}_{\kern-1ptF}(y,x'; g_y, g_x) \geq \oset{D}_{\kern-1ptF}(y,x; g_y, g_x)$ if $x' \geq x$, and nonincreasing in outputs, i.e., if $y' \leq y$ and $\oset{D}_{\kern-1ptF}(y',x; g_y, g_x) > -\infty$, then $\oset{D}_{\kern-1ptF}(y',x; g_y, g_x) \geq \oset{D}_{\kern-1ptF}(y,x; g_y, g_x)$;
		\item[D6.]  $\oset{D}_{\kern-1ptF}$ is a proper concave function of $(y,x)$.
	\end{enumerate}
	\textsc{Proof.} Properties D1 and D2 follow from (3.2.1) and Proposition 3.1.2, property D3 follows from F1 and F3, whereas property D6 follows from F1--F4.
	
	Next, we show that $\oset{D}_{\kern-1ptF}$ satisfies representation property D4 whenever $F$ satisfies properties F2 and F3. The general version of D4 was proved by Chambers et al.\ (1998, pp.\ 354--355). 
	
	Fix $(y,x) \in \rp$ and $(g_y, g_x) \in \rp\setminus\{(0,0)\}$. First, suppose that $F(y,x) \leq 0$. Since $0 \in \Gamma$ and $F_{[S]}(0) \leq 0$, it follows that $0 \leq \sup \Lambda = \oset{D}_{\kern-1ptF}(y,x; g_y, g_x)$. Conversely, suppose $\oset{D}_{\kern-1ptF}(y,x; g_y, g_x) \geq 0$. Then the set $\Lambda$ is nonempty and either $\Lambda = \Gamma$ or $\Lambda \subset \Gamma$. If the sets $\Lambda$ and $\Gamma$ are equal, then $F_{[S]}(\beta) \leq 0$ for all $\beta \in \Gamma$, and consequently $F_{[S]}(0) \leq 0$. If $\Lambda$ is a proper subset of $\Gamma$, then $G_S(0) \geq 0$ by Lemma 3.2.3. In this case, $0 \in$ ran $F_{[S]}$, and therefore $G_S(0) = \beta^\ast$ implies $F_{[S]}(\beta^\ast) = 0$. From this result and strict monotonicity of $F_{[S]}$, it follows that $F_{[S]}(0) \leq 0$.
	
	It remains to prove that $\oset{D}_{\kern-1ptF}$ satisfies property D5. The DTDF is nondecreasing in inputs, since $\Lambda(y,x; g_y, g_x) \subseteq \Lambda(y,x'; g_y, g_x)$ if $x' \geq x$, whenever property F3 holds. Next, suppose $y' \leq y$ and $\oset{D}_{\kern-1ptF}(y',x; g_y, g_x) > -\infty$. We prove only the case when $\Lambda(y,x; g_y, g_x)$ is a nonempty proper subset of $\Gamma(y,x; g_y, g_x)$. Let $\beta^\ast = \oset{D}_{\kern-1ptF}(y,x; g_y, g_x)$ and suppose, first, that $\beta^\ast \in\Gamma(y',x; g_y, g_x)$. Property F3 implies that $F(y' + \beta^\ast g_y, x - \beta^\ast g_x) < 0$, and therefore $\beta^\ast \leq \sup\Lambda(y',x; g_y, g_x)$. Next, suppose that $\beta^\ast \not\in \Gamma(y',x;g_y,g_x)$. Then
	$$\beta^\ast < \max\{-y_j'/g_{yj}\mid j \in \mathrm{J}^{+}(g_y)\} \leq \beta \leq \sup\Lambda(y',x; g_y, g_x)$$
	for every $\beta \in \Lambda(y',x; g_y, g_x)$, which is nonempty by assumption.  $\qed$ 
	
	Finally, it follows from property D4 that, for all $(g_y,g_x) \in \rp\setminus\{(0,0)\}$, the sets $\{(y,x) \in \rp\mid \oset{D}_{\kern-1ptF}(y,x; g_y, g_x) \geq 0\}$ and $\{(y,x) \in \rp\mid F(y,x) \leq 0\}$ are equal, and both of them satisfy properties T1--T7 by Theorem 2.2.1.
	
	For expository purposes, we end this section with an example of how the directional technology distance function can be explicitly derived from a symmetric transformation function satisfying properties F1--F4.
	
	\textbf{Example 3.2.7.} Consider the transformation function $F$, separable in inputs and outputs\footnote{See, for instance, Lau (1972, p.\ 284), Hall (1973), and Chambers and Färe (1993) for a discussion of separability.}, that is given by $F(y,x) = q(y) - f(x)$, where $q\colon \ro \to \R$ is a quadratic output function defined by $q(y) = \sum_{k=1}^{m}\beta_k y_k + \frac{1}{2}\sum_{k = 1}^{m}\sum_{\ell=1}^{m}\beta_{k\ell}y_k y_\ell$, with $\beta_{k\ell} = \beta_{\ell k}$ for all $k,\ell \in \{1, \dots, m\}$, and $f\colon \ri \to \R$ is a linear input function defined by $f(x) = \sum_{i=1}^{n}\alpha_i x_i$. Let $b$ denote a vector in $\R^m$ with the $k$th component $\beta_k$ and $a$ denote a vector in $\R^n$ with the $i$th component $\alpha_i$. Also, let $B$ denote an $m$-by-$m$ symmetric matrix $\bigl[\beta_{k\ell}\bigr]$ and write the elements of $\R^n$ or $\R^m$ in column-vector form. Then
	\begin{equation*}
	F(y,x) =  \trans{b}y + \tfrac 12\trans{y}By - \trans{a}x, \qquad B = \trans{B}.
	\end{equation*}
	If its parameters satisfy the following constraints: $b > 0$, $a > 0$, and $B$ is a nonnegative and positive semidefinite matrix, then $F$ satisfies properties F3 and F4.
	
	Fix $(y,x) \in \rp$ and $(g_y, g_x) \in \rp\setminus\{(0,0)\}$. The restriction of $F$ to the line segment or the ray $S$ is given by
	\begin{equation*}
	F_{[S]}(\beta) = \beta^2(\tfrac 12\trans{g_y}B g_y) + \beta(\trans{b}g_y + \trans{y}B g_y + \trans{a} g_x) + (\trans{b}y + \tfrac{1}{2}\trans{y}B y - \trans{a}x),
	\end{equation*} 
	where the constant term equals $F(y,x)$ and the quadratic term vanishes if $\trans{g_y}B g_y = 0$. The directional technology distance function $\oset{D}_{\kern-1ptF}$ is then given by (3.2.5), with
	\begin{equation*} 
	G_S(0) = [\trans{g_y}B g_y]^{-1}[-(\trans{b}g_y + \trans{y}B g_y + \trans{a}g_x) + ((\trans{b}g_y + \trans{y}B g_y + \trans{a}g_x)^2 - 2(\trans{g_y}Bg_y)F(y,x))^{1/2}]
	\end{equation*}
	if $\trans{g_y}Bg_y > 0$, and $G_S(0) = -F(y,x)/(\trans{b}g_y + \trans{a}g_x)$ if $\trans{g_y}Bg_y = 0$, and satisfies properties D1--D6 by Theorem 3.2.6. Figure 4 illustrates the case when $b = (1,1)$, $a = (1,1)$, $B = \text{diag}(1,1)$, $y = (0.5,0.5)$, $x = (1,1)$, $g_y = (0.5,0.5)$, and $g_x = (0,0)$. 
	
	\begin{figure}[H]
	\centering
	\begin{minipage}[b]{0.48\textwidth}
		\includegraphics[width=\textwidth,trim={1.5cm 1.5cm 1.5cm 1.5cm},clip]{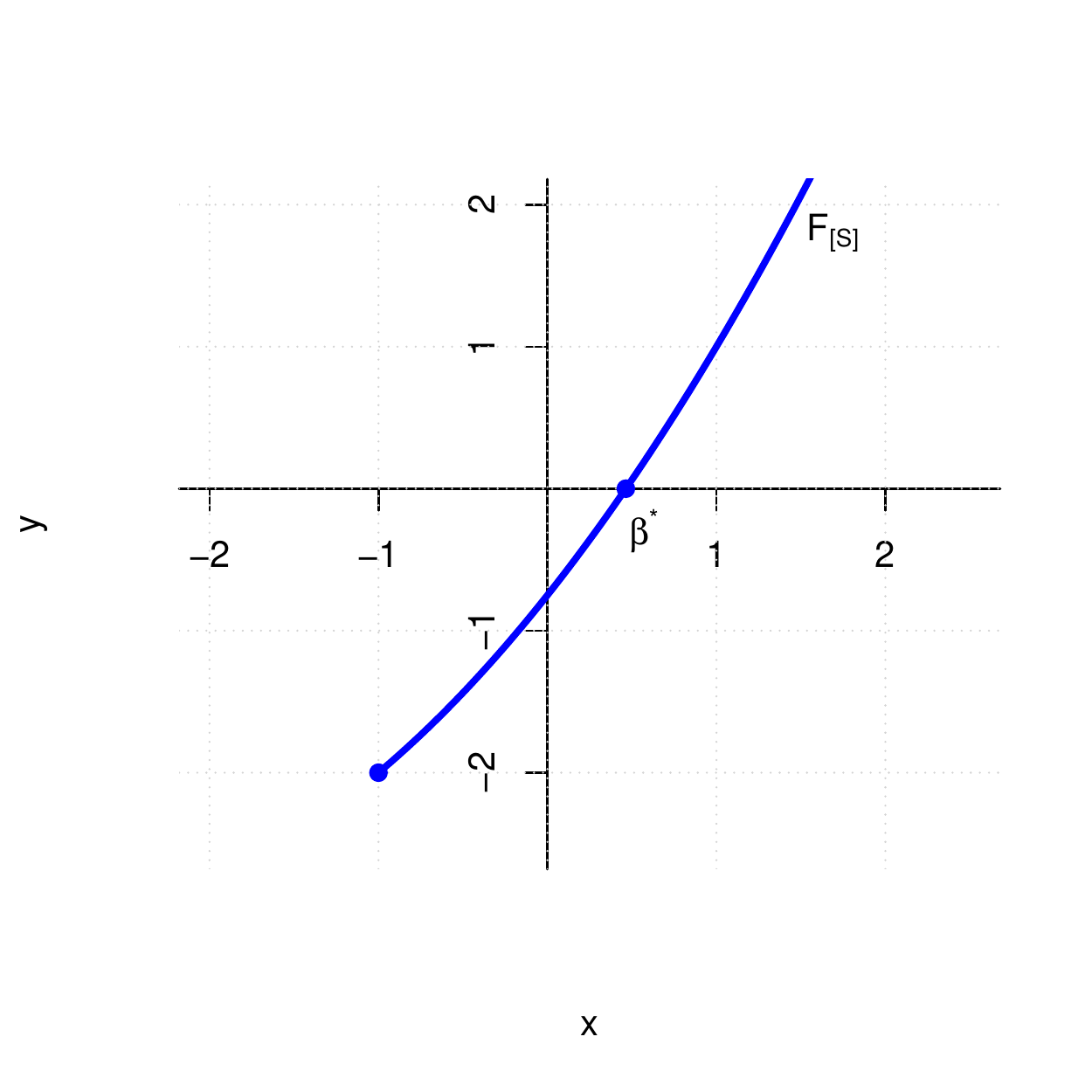}
	\end{minipage}
	\hfill
	\begin{minipage}[b]{0.5\textwidth}
		\includegraphics[width=\textwidth,trim={0 0.5cm 1cm 1.5cm},clip]{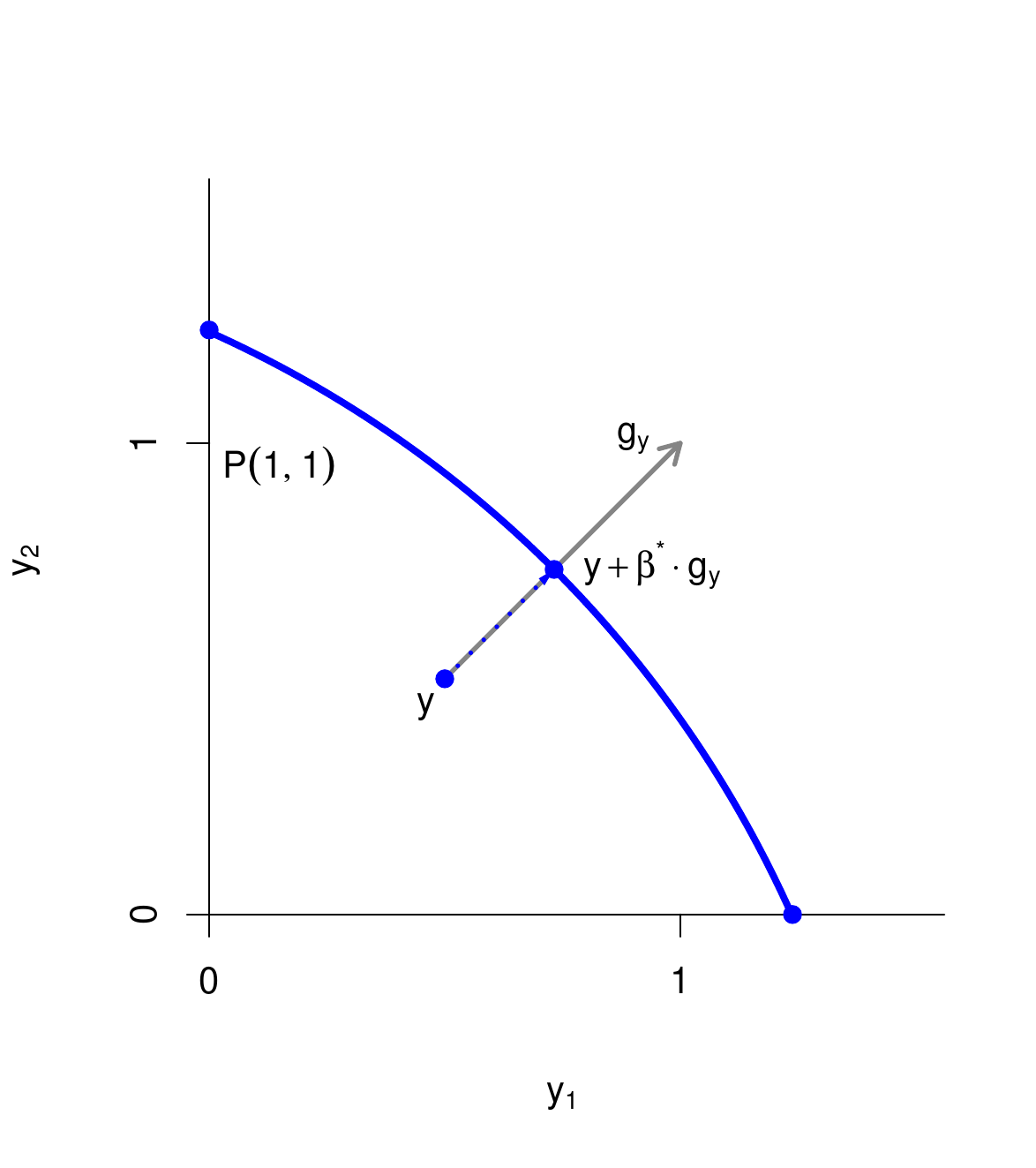}
	\end{minipage}
		\caption{$F_{[S]}$ is strictly increasing and convex on $[-1, \infty)$ and assumes the value zero at $\beta^\ast \approx 0.464$.}
	\end{figure}

	The approach suggested in this section has certain limitations. A number of functional forms for input, output, and nonseparable transformation functions have been discussed in the literature, and various methods have been used to ensure that these functions satisfy appropriate monotonicity and convexity or concavity conditions; see, for example, Diewert (1973, 1974), Hasenkamp (1976), Lau (1978), Diewert and Wales (1987), and Kumbhakar (2011). However, for most of these functional forms, the zero of $F_{[S]}$ cannot be expressed in closed form, and therefore numerical approximation techniques have to be employed. Further research is needed to determine whether or not this is feasible in the context of econometric estimation of the systems of simultaneous equations including the directional technology distance function. 
	\newpage

	\section*{Appendix}
	For convenience, we repeat here the definitions of the isoquant, the weak efficient, and the efficient subsets of an input and output sets from Färe et al.\ (1985, p.\ 28) and Färe et al.\ (1994, pp.\ 39--40). 
	
	\textbf{Definition 1.} The isoquant of an input set is defined as
	\begin{equation*}
	\text{Isoq } L(y) = \{x \in L(y)\mid \lambda x \not\in L(y) \text{ if } \lambda \in [0,1)\}
	\end{equation*}
	if $y \in Y_1$, and Isoq $L(0) = \{0\}$.
	
	\textbf{Definition 2.} The weak efficient subset of an input set is defined as
 	\begin{equation*}
 	\text{WEff } L(y) = \{x \in L(y)\mid x' \not\in L(y) \text{ if } x' \stackrel{\ast}{<} x\}
 	\end{equation*}
 	if $y \in Y_1$, and WEff $L(0) = \{0\}$.
 	
 	\textbf{Definition 3.} The efficient subset of an input set is defined as
 	\begin{equation*}
 	\text{Eff } L(y) = \{x \in L(y)\mid x' \not\in L(y) \text{ if } x' \leq x\}
 	\end{equation*}
 	if $y \in Y_1$, and Eff $L(0) = \{0\}$. 
 	
	\textbf{Definition 4.} The isoquant of an output set is defined as
	\begin{equation*}
	\text{Isoq } P(x) = \{y\in P(x)\mid \theta y \not\in P(x) \text{ if } \theta > 1\}
	\end{equation*}
	if $x \in X_1$, and Isoq $P(x) = \{0\}$ if $x \in X_2\cup X_3$.
	
	\textbf{Definition 5.} The weak efficient subset of an output set is defined as
	\begin{equation*}
	\text{WEff } P(x) = \{y\in P(x)\mid y' \not\in P(x) \text{ if } y' \stackrel{\ast}{>} y\}
	\end{equation*}
	if $x \in X_1$, and WEff $P(x) = \{0\}$ if $x \in X_2\cup X_3$.
	
	\textbf{Definition 6.} The efficient subset of an output set is defined as
	\begin{equation*}
	\text{Eff } P(x) = \{y\in P(x)\mid y' \not\in P(x) \text{ if } y' \geq y\}
	\end{equation*}
	if $x \in X_1$, and Eff $P(x) = \{0\}$ if $x \in X_2\cup X_3$.
	\newpage
	
	\section*{Acknowledgments}
	Part of this work was carried out during the research stay at the State University of New York at Binghamton in the spring of 2017. I am greatly indebted to Subal C. Kumbhakar for many insightful discussions. I am also sincerely grateful to Oleg Badunenko, Karl Mosler, Sven Otto, Andriy Regeta, and Dominik Wied for helpful comments and suggestions. This research was partially supported by a PhD scholarship from the Cologne Graduate School in Management, Economics and Social Sciences at the University of Cologne.

\end{document}